\documentclass[bibnotes,floats,aps,a4paper,twocolumn,showpacs,superscriptaddress]{revtex4}
\usepackage{epsfig}
\usepackage{times} 
\usepackage{graphicx} 
\usepackage{amsmath}
\begin{document}

\title{  
Dewetting of thin films on heterogeneous substrates: 
Pinning vs. coarsening 
}
\author{Lutz Brusch}
\author{Heiko K\"uhne}
\affiliation{Max-Planck-Institut f\"ur Physik komplexer
  Systeme, N\"othnitzer Stra{\ss}e 38, D-01187 Dresden, Germany}
\email{baer@mpipks-dresden.mpg.de}
\author{Uwe Thiele}
\affiliation{Max-Planck-Institut f\"ur Physik komplexer
  Systeme, N\"othnitzer Stra{\ss}e 38, D-01187 Dresden, Germany}
\affiliation{Department of Physics, 
University of California, Berkeley, CA 94720-7300, USA}
\email{thiele@mpipks-dresden.mpg.de}
\author{Markus B\"ar}
\affiliation{Max-Planck-Institut f\"ur Physik komplexer
  Systeme, N\"othnitzer Stra{\ss}e 38, D-01187 Dresden, Germany}

\date{\today} 

\begin{abstract}

We study a model for a thin liquid film dewetting 
from a periodic heterogeneous substrate (template). 
The amplitude and periodicity of a striped template
heterogeneity necessary to obtain a stable periodic 
stripe pattern, {\it i. e.} pinning, are computed. 
This requires a stabilization of 
the longitudinal and transversal 
modes driving the typical coarsening dynamics during
dewetting of a thin film on a homogeneous substrate. 
If the heterogeneity has a larger spatial period 
than the critical dewetting mode, 
weak heterogeneities are sufficient for pinning.
A large region of coexistence between 
coarsening dynamics and pinning is found. 

\end{abstract}

\pacs{
68.15.+e, 
68.55.-a, 
47.20.Ma, 
47.20.Ky  
}
\maketitle


Templating and controlled rupture of liquid films on chemically structured 
substrates have provoked many experimental efforts 
\cite{Rock99,MKPE99,GHLL99,KT99,GCF00,HJ00} but
the conditions for the desired imaging of the template structure onto  
the deposit pose many open questions. 
First theoretical results for liquid layers on
structured surfaces with strong heterogeneities of wettability, {\it i.e.,} 
stepwise alternating hydrophilic and hydrophobic stripes in an aqueous system, 
describe different morphological transitions when changing the size of the
heterogeneous patches or the volume of the deposited liquid 
\cite{LeLi98,BaDi00,KKS00,KaSh01}.
Experimentally, the challenge of preventing the dewetting pattern from 
coarsening has been met by evaporation of the solvent \cite{TMP98}, 
by literally freezing the system \cite{BSHL96} or by using a heterogeneous 
substrate \cite{Rock99}.

In the present paper we study the transition between coarsening and pinning 
for thin films on weakly structured substrates that
possess spatial modulations of the  
molecular interaction terms.
Thin films on homogeneous substrates can be unstable to spinodal
dewetting, see {\it e.g.} \cite{Re92,He98}. 
Sharma and coworkers proposed a model that contains polar and apolar
components of molecular interactions and reproduced the dynamics of
dewetting thin films \cite{KhSh98}. 
Recently, it has been shown that this model 
has periodic stripe solutions that are unstable to coarsening 
\cite{TVN01}.
Here we use a slightly different model derived from diffuse interface 
theory by Pismen and Pomeau \cite{PiPo00} as a starting point. 
We emphasize that this model has very similar dynamics \cite{BeNe01}
and stationary solutions \cite{TVNP01} as the Sharma model. 
The impact of a heterogeneous substrate on the stationary film
profiles, their linear stability and 
resulting conditions for morphological transitions are presented. 
The dynamics of thin films on periodic stripe templates have been
studied in \cite{KaSh01} with the original Sharma model by numerical 
integration and for a given heterogeneity of large amplitude. 
Using bifurcation and stability analysis, we can provide, however, 
a more systematic study of the effects of parameters like average
film thickness $\bar{h}$, amplitude $\epsilon$ and spatial period 
$P_{het}$ of the heterogeneity. 

\begin{figure}\vspace{-1mm}
\begin{center}
 \epsfxsize=1.\hsize \mbox{\hspace*{-.06 \hsize} \epsffile{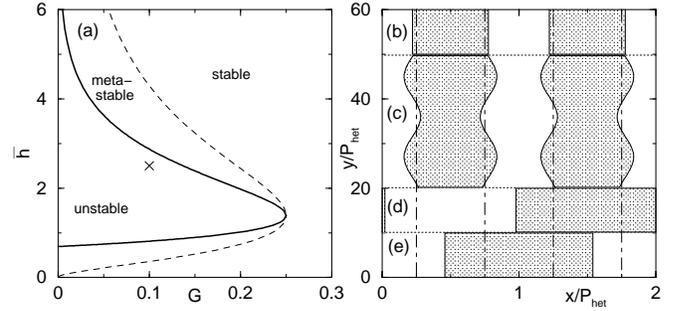} }
\end{center}
 \vspace{-.5cm}
 \caption
 {(a) Phase diagram for thin films on a homogeneous substrate 
 (after \cite{TVNP01}).
  The cross shows the parameter values for which we present results 
  in detail.
  (b) Schematic display of the template (dot-dashed line), the
  periodic film profile with the same spatial period (solid line).
  Shaded areas show where $h>\bar{h}$. 
  (c) Initial stage of the transversal instability and (d),(e) final stages 
  of variants of the longitudinal coarsening instability. 
 }
 \vspace{-.5cm}
 \label{scheme}
\end{figure}

Our starting point is a homogeneous or weakly heterogeneous substrate. 
We choose parameters in the unstable regime (cross in Fig.~\ref{scheme}(a)). 
Then a striped solution of a spatial period $P=P_{het}$ is considered. 
For illustration we display a system of length $2 P_{het}$ (see
Fig.~\ref{scheme}(b)-(e)). 
The periodic stripe pattern (Fig.~\ref{scheme}(b)) is now unstable against several 
transversal (Fig.~\ref{scheme}(c)) and longitudinal 
(Figs.~\ref{scheme}(d),(e)) perturbations. 
Stability analysis allows us to trace the corresponding eigenvalues. 
Increasing the strength of the heterogeneity, 
they all will acquire negative real
parts and the desired pattern in Fig.~\ref{scheme}(b) is pinned, 
{\it i.e.} becomes stable against small perturbations. 

The evolution equation for the film thickness profile $h(x,y,t)$
contains a disjoining pressure that is derived by combining the Stokes equation 
in long wave approximation \cite{ODB97}
with diffuse interface theory \cite{PiPo00}: 
\begin{equation}
\partial_t h  =  -\nabla \{ (h-\ln a)^3 \nabla [\Delta h 
 - \partial_h f(h,x) ] \}
 \label{h4}
\end{equation}
with a free energy 
$f(h,x) = \kappa(x)\, e^{-h} (e^{-h}-2)\,+\, \frac{1}{2}G h^2$ 
and the ratio $G$ 
of gravitation to mean molecular interactions as well as
the spatially varying strength of the molecular interactions $\kappa(x)$.
$a>0$ is a small parameter describing the wetting properties in 
the regime of partial wetting \cite{PiPo00,TVNP01}.
This dimensionless form is obtained after scaling the original quantities
as in Ref.~\cite{TVNP01}.
%
%
The molecular interactions that become important on the nanometer scale 
\cite{Isra92} are incorporated through the disjoining pressure $\Pi(h)$ 
contained in the derivative of the free energy
$\partial_h f(h,x)=-\kappa(x) \Pi(h)+ G h$.
The chosen $\Pi(h)$
is qualitatively equivalent to other pressures consisting of destabilizing 
short-range and stabilizing long-range interactions \cite{ODB97} but does not 
suffer from the usual divergency for vanishing film thickness \cite{TVNP01}. 
In the absence of heterogeneity, the model possesses two control
parameters, the dimensionless quantity $G$ and the average film
thickness $\bar{h}$ that is chosen as conserved quantity. 
A phase diagram in $G$ and $\bar{h}$ indicating the region of
spinodal dewetting is shown in Fig.~\ref{scheme}(a).  

Heterogeneous substrates with a smooth change in the wettability are
modelled by a spatial sinusoidal modulation of the overall strength of 
the disjoining pressure
\begin{equation}
 \kappa(x) = 1+\epsilon \cos (2\pi x/P_{het})~.
 \label{h3}
\end{equation}
The system behaviour is controlled by the amplitude, $\epsilon$, 
and the imposed periodicity, $P_{het}$ of the heterogeneity.
Locally small $\kappa$ correspond to effectively larger $G$ 
\cite{TVNP01}.
Droplets are preferably located at minima of $\kappa(x)$.

\begin{figure}\vspace{-1mm}
\begin{center}
 \epsfxsize=1.\hsize \mbox{\hspace*{-.06 \hsize} \epsffile{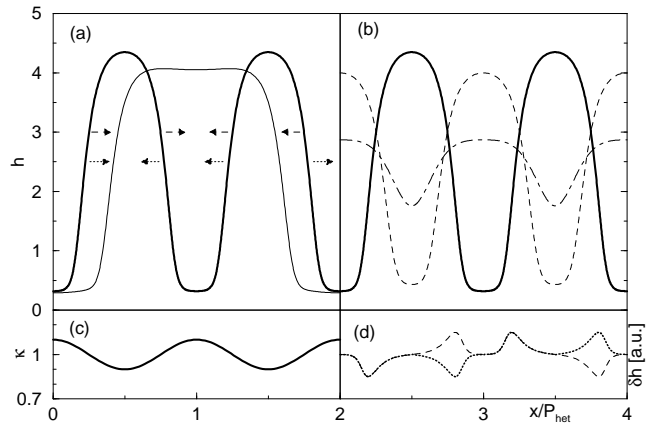} }
\end{center}
 \vspace{-.5cm}
 \caption
 {(a) Stationary film profiles to Eq.~\ref{h4} with $P=P_{het}$ (thick) and $P=2
 P_{het}$ (thin).
 (b) Portraits of all solutions with $P=P_{het}$. Only the solid profile is
 linearly stable.
 (c) The heterogeneity and (d) the two coarsening modes due to translation
 (dashed) and transfer of mass (dotted) as sketched in (a).
 Parameters are $\epsilon=0.1, G=0.1, \bar{h}=2.5, P_{het}=50$.
 }
 \vspace{-.5cm}
 \label{ampl}
\end{figure}

We choose the system size, $L$, as a multiple of $P_{het}$,
apply periodic boundary conditions and vary $\epsilon \ge 0$.
The Lyapunov functional
\begin{equation}
E  =  \frac{1}{L} 
     \int_0^L\left[\frac{1}{2}(\partial_x h)^2 + f(h,x)\right]dx
 \label{h7}
\end{equation}
gives the energy of a striped solution $h(x,t)$ \cite{TVNP01}.
During the time evolution of a given initial film thickness profile this
energy decreases and eventually settles in a minimum when the system approaches
a linearly stable stationary solution of Eq.~(\ref{h4}). 
One can determine these solutions directly
by setting $\partial_t h=0$ in Eq.\,(\ref{h4}) and integrating twice
\cite{TVNP01}, yielding
\begin{equation}
0=\partial_{xx} h - \partial_h f(h,x) + C_1~.
 \label{h9}
\end{equation}
%
The integration constant $C_1$ and the spatial period $P$ now parametrize 
a two-parameter family of periodic solutions, 
{\it i.e.} periodic patterns of holes and droplets, that can be calculated
by continuation techniques \cite{AUTO97}. $P$ can be equal to $P_{het}$ or its 
multiples. 
Focusing on situations with conserved liquid volume we adopt
the mean film thickness $\bar{h}=1/L \int_0^L h(x) dx$ as the second 
parameter beside the period, using $C_1$ as a Lagrange multiplier.
Typical film profiles are shown in Fig.~\ref{ampl}(a),(b).

For homogeneous substrates with $\epsilon=0$,
flat films are stationary solutions that, depending
on mean film thickness and the parameter $G$, may be unstable to
infinitely small (spinodal dewetting) or finite (nucleation) disturbances of the
film surface. 
Rupture due to the spinodal mechanism occurs for
perturbations with periods larger than a critical value, 
$P_c=2\pi/k_c$ with $k_c^2=-\partial_{hh} f(h,x)|_{\bar{h}}$  \cite{TVNP01}.
On a slower time scale the initially formed holes coalesce, the 
pattern becomes coarser and tends to the absolute minimum of the energy at the 
largest possible $P$, {\it i.e.} the system size.

\begin{figure}\vspace{-1mm}
\begin{center}
 \epsfxsize=1.\hsize \mbox{\hspace*{-.06 \hsize} \epsffile{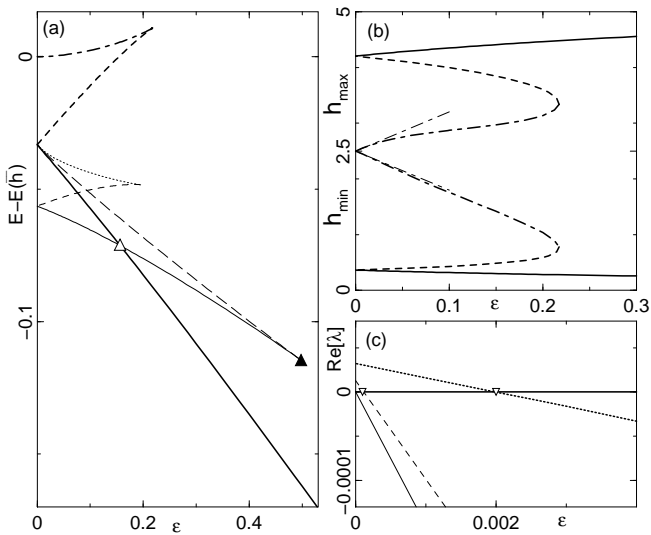} }
\end{center}
\vspace{-.5cm}
 \caption
 {(a) Relative energy of stationary solutions to Eq.~(\ref{h4}) with 
 $P=P_{het}$ (thick) and $P=2 P_{het}$ (thin) versus $\epsilon$. 
 A flat film at $\epsilon\!=\!0$ yields $E(\bar{h})\!\approx\!0.155$.
 Solid curves correspond to coexisting linearly stable solutions with.
 (b) Bifurcation diagram representing maximum (upper half) and minimum (lower
 half) of the solutions with $P=P_{het}$. 
 The approximation (\ref{ana2}) is denoted by
 the thin dot-dashed lines and other line styles match with (a) and
 Fig.~\ref{ampl}(b).
 (c) Eigenvalues $\lambda$ with largest real part of the lowest energy 
 solutions with $P=P_{het}$ (thick solid branch in (a)).
 Solid curves correspond to Goldstone modes at $\epsilon=0$ and 
 broken curves to interaction modes with period $2 P$. 
 Line styles match the profiles in Fig.~\ref{ampl}(d).
 Triangles correspond to period doubling bifurcations where solutions with $P=2
 P_{het}$ emerge.
 Parameters are $a=0.1, G=0.1, \bar{h}=2.5, P_{het}=50$ which yields
 $\alpha/\epsilon\approx 0.7$.
 }
 \vspace{-.5cm}
 \label{eps1}
\end{figure}

{\it (i) Perturbation theory  -}
If the heterogeneity is switched on, the flat film is no longer a
solution to Eq.~(\ref{h9}). However, for very small $\epsilon\ll1$ 
an analytical expression for the weakly varying stationary film profile 
can be calculated. Rewriting the heterogeneity as 
$\epsilon (e^{i k_{het} x}+c.c.)/2$ with $k_{het}=2\pi/P_{het}$ and using the volume 
conserving ansatz $h(x)=\bar{h}+\alpha (e^{i k_{het} x}+c.c.)/2$ with 
$O(\alpha)=O(\epsilon)$ in 
Eq.~(\ref{h9}), gives to first order in $\epsilon$ a
stationary solution with 
\begin{equation}
 \alpha = \frac{-\Pi(\bar{h})}{k_c^2-k_{het}^2} ~ \epsilon~.
   \label{ana2} 
\end{equation}
This solution is only valid for $\alpha\ll1$.
For $P_c < P_{het}$, the solution is modulated in phase with
the heterogeneity, whereas for $P_c > P_{het}$ the phase is
shifted by $\pi$. 
Results obtained by perturbation theory are depicted in Fig.~\ref{eps1}(b). 


{\it (ii) Numerical bifurcation results  -}
Suppose a variety of heterogeneous substrates with appropriate period $P_{het}$ but
different $\epsilon$ is available and a dewetting pattern with spatial period 
$P=P_{het}$ is desired. 
We choose a linearly unstable mean film thickness $\bar{h}$ such that the critical
wavelength of spinodal dewetting $P_c$ is of the same order as
$P_{het}$.
As an example, we  choose  $P_c\!\approx\! P_{het}/1.5$
with $P_{het}\!=\!50$ and $G\!=\!0.1, \bar{h}\!=\!2.5$. 
This ratio of $P_{het}/ P_c$  is known to give good templating for large
heterogeneities \cite{KaSh01}.
The stationary solutions with $P=P_{het}$ and $P=2 P_{het}$ are computed 
\cite{noteallsol} as 
$\epsilon$ is increased using the weakly modulated solutions as a starting 
point for the continuation \cite{AUTO97}.
Figs.~\ref{ampl}(a),(b) and \ref{eps1}(a) show profiles of the 
solutions and the bifurcation diagram 
and Fig.~\ref{eps1}(b) compares the results with the small $\epsilon$ 
approximation Eq.\,(\ref{ana2}).
The weakly modulated solutions are indeed in line with the results of
perturbation theory. 
Breaking the translational symmetry at $\epsilon=0$ gives rise to two branches 
of solutions starting at the large amplitude solution known from the
homogeneous case \cite{TVNP01}. 
Only the branch of largest amplitude possesses a phase shift. 
The branches of solutions in phase cease to exist in a saddle-node bifurcation
at $\epsilon\!\approx\! 0.22$.


{\it (iii) Energy and longitudinal stability -}
Thick curves in Fig.~\ref{eps1}(a) compare the energies 
(calculated with Eq.~(\ref{h7})) of the different solution branches
given in Fig.~\ref{eps1}(b). 
For the chosen parameter values 
the solutions in phase have always a larger energy than the ones out of phase.
When increasing $\epsilon$ the energy of the lower branch decreases further 
indicating that the heterogeneity favors this pattern with $P=P_{het}$. 
Thin curves in Fig.~\ref{eps1}(a) denote solutions with $P=2 P_{het}$. 
%
To gain more insight into the conditions of pinning the pattern
by a heterogeneity, we analyse the linear stability of the 
stationary solutions in systems of different sizes \cite{stab}.
For the stability analysis of periodic solutions, one usually 
employs a Floquet ansatz. 
This enables us to get the stability of periodic solutions in large
systems corresponding to large ratios $n = L/P_{het}$. 
We find however that the most dangerous, potentially unstable
longitudinal modes that induce coarsening of dewetting films 
are already present in systems for $n=2$. 
Hence, we constrain the discussion to system sizes $L \le 2
P_{het}$ in $x-$direction.
In the transversal $y$-direction, we had to use much larger 
length $\approx 30 P_{het}$ for a stability analysis, see below.

In a system of size equal to one period of the heterogeneity, {\it i.e.}
$L=P_{het}$, the entire lower branch in Fig.~\ref{eps1}(a) is
linearly stable, whereas the two other branches are unstable.
For $\epsilon=0$ the linearly stable solution possesses two zero eigenvalues 
(Goldstone modes) corresponding to translation symmetry and mass
conservation. 
For $\epsilon\neq0$ the translation symmetry is 
broken and the corresponding eigenvalue becomes negative 
(thin solid line in Fig.~\ref{eps1}(c)).
Mass conservation is maintained and  the corresponding
eigenvalue remains zero for all $\epsilon$ 
(thick solid line in Fig.~\ref{eps1}(c)).

In a larger system with $L=2 P_{het}$, 
the first stage of coarsening can be studied which is the fastest since nearest 
neighbours interact and the energy gain is largest. 
If templating shall be
successful then this first stage needs to be prevented by pinning. 
In the 
stability analysis of the two periods, two new eigenvalues appear that 
correspond to asymmetric combinations of the Goldstone modes 
(see Fig.~\ref{ampl}(d)). 
They represent two possible modes of coarsening:
(a) Shift of droplets towards each other 
caused by a combination of opposite translational modes
(dashed arrows in Fig.~\ref{ampl}(a) and dashed curves in Figs.~\ref{ampl}(d) 
and \ref{eps1}(c));
(b) Mass transfer between neighbouring unmoved droplets caused by a combination of 
opposite volume modes (dotted arrows in Fig.~\ref{ampl}(a) and 
dotted curves in Figs.~\ref{ampl}(d) and \ref{eps1}(c)).
These modes have first been recognized in the Cahn-Hilliard equation
\cite{Kawa82}. 
For small $\epsilon$ both eigenvalues are positive, 
implying the instability of the wanted pattern to coarsening.
The mass transfer proceeds faster than the shift of droplets. 
But, as $\epsilon$ increases, both become negative, rendering the 
solution with period $P_{het}$ linearly stable. 
At the two crossing points
period doubling bifurcations occur where stationary
solutions of period $P=2 P_{het}$ emerge (compare thin branches 
in Fig.~\ref{eps1}(a)).
Altogether, the desired pinning solution with $P=P_{het}$ is linearly stable
for $\epsilon>0.002$.

The value of this critical $\epsilon$ depends
on parameters and the specific form of $\Pi(h)$.
Since the two symmetries connected with the coarsening modes
are present for arbitrary choices of $\Pi(h)$ the result does not depend on 
this choice. 
Accordingly, two pairs of solutions with $P=2 P_{het}$ 
appear in any given system at finite $\epsilon$.

These four branches of solutions with $P=2 P_{het}$ (thin curves in
Fig.~\ref{eps1}(a)) have the following linear stability.
Solutions on the  dotted branch carry two positive eigenvalues. 
%
The short-dashed branch has still one positive eigenvalue that leads to a shift
of the pattern towards solutions of the solid branch.
The long-dashed branch has one positive eigenvalue and is a saddle that
divides evolutions by translation of two droplets towards one on the solid
branch or back to the $P=P_{het}$ branch.
The entire thin solid branch is linearly stable in $L=2 P_{het}$ and
represents the coarse solution competing with the desired pattern (see
Fig.~\ref{scheme}(e) {\it resp.} the thin profile in Fig.~\ref{ampl}(a)).
Here, at large $\bar{h}$, the coarse profile of lowest energy emerges from the 
translation mode, whereas for smaller mean film thicknesses 
($\bar{h}\le 2$ for $G=0.1$) the coarse profile corresponding to the transfer
mode is stable (see Fig.~\ref{scheme}(d)) and has lowest energy \cite{tobepubl}.


{\it (iv) Coarsening versus pinning -}
As seen in {\it (iii)} coarsening is favored for $\epsilon<0.002$ 
and solutions with $P=2 P_{het}$ are the only stable ones. 
Consequently they also have the lowest energy (for $L=2 P_{het}$).
For $\epsilon>0.002$ multistability between linearly stable 
solutions with $P=P_{het}$ and
$P=2 P_{het}$ occurs and the initial condition becomes crucial for the final
dewetting structure. 
Below $\epsilon\approx0.157$ the coarse solution
has lowest energy, whereas at larger $\epsilon$ the desired pinned pattern 
is energetically favored. 
%
%
Choosing suitable initial conditions enables pinning at much
smaller $\epsilon$. 
But coarsening will still occur at larger $\epsilon$
if initial conditions are chosen accordingly. 
At even larger $\epsilon>0.5$ the pinned pattern is the only
possibility  because the coarse solutions cease to exist.

\begin{figure}\vspace{-1mm}
\begin{center}
 \epsfxsize=1.\hsize \mbox{\hspace*{-.06 \hsize} \epsffile{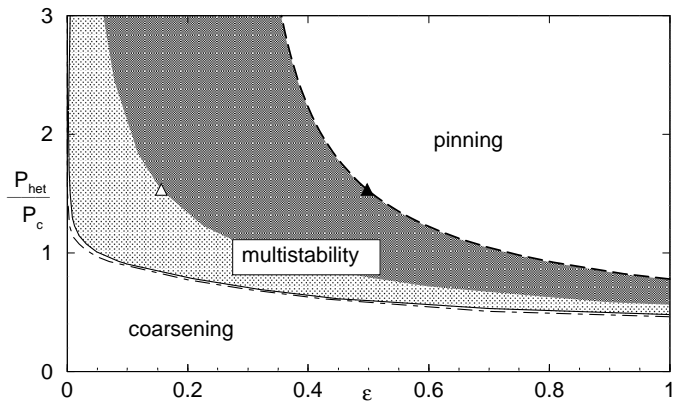} }
\end{center}
 \vspace{-.5cm}
 \caption
 {Morphological phase diagram of templating showing regions in the parameter 
 plane $(\epsilon, P_{het})$ 
 with different behavior of the thin film on a heterogeneous substrate.
 The shaded band separates parameters of ultimate coarsening from ultimate
 pinning. Inside the shaded band multistability is present with the desired
 pattern being the energetically minimum inside the dark shaded
 area. Parameters are $a=0.1, G=0.1, \bar{h}=2.5, P_c\approx33$.
 The triangles correspond to the equivalent symbols in Fig.~\ref{eps1}(a). 
 }
 \vspace{-.5cm}
 \label{phase}
\end{figure}

Summarizing the results for a range of values $P_{het}$ at fixed $G=0.1$ and
$\bar{h}=2.5$ we obtain the
``morphological phase diagram'' Fig.~\ref{phase}. 
The product $\epsilon*P_{het}$ emerges as a rough fundamental parameter:
Coarsening prevails when it has low values while for large values 
the pattern pins to the heterogeneity as desired.
At intermediate values  multistability is found where
the initial condition selects the final outcome. 


{\it (v) Transversal stability -}
The transversal stability of the pinned profiles ($P=P_{het}$)
can be obtained from the
eigenvalues $\lambda$ of small perturbations $\delta h(x) \exp (\lambda t)
(\exp (i k y) + c.c.)$~. 
We observe a long wavelength instability at small
$\epsilon$ on a length scale much larger than the longitudinal modes.
The most dangerous branch of transversal perturbations emerges from
the zero-eigenvalue representing mass conservation in the film.
For $\epsilon = 0.001$ for instance, we found 
that the fastest growing mode had a wavelength of $\approx 900 = 18
P_{het}$ (see Fig.~\ref{scheme}(c)). 
The transversal modes are stabilized for parameters above the solid 
curve in Fig.~\ref{phase}. 
For the present choice of $\Pi(h)$ and $G, \bar{h}$ 
the transversal instability
restricts the stability range of the pinned solution more than
longitudinal coarsening.


{\it (vi) Summary -}
We have studied the conditions for successful templating or pinning of thin
films that are unstable to spinodal dewetting. 
The main result is that a periodic stripe pattern can be pinned if
the heterogeneous substrate suppresses the transverse and longitudinal
instabilities experienced by such patterns on a homogeneous
substrate. 
While previous studies have employed direct numerical simulation to
study the dynamics of thin films, we have used numerical bifurcation
and stability computations. 
This allows an efficient scanning of parameters characterizing the
heterogeneous substrate and the film dynamics. 
Comparing the two length scales $P_c$ and $P_{het}$ of spinodal dewetting and
heterogeneity, respectively, we find 
that patterns are not pinned to heterogeneities on a 
much smaller scale than $P_c$, and  that smaller $P_c$ need weaker 
heterogeneities to pin the pattern.
In consequence, templating can be best controlled by choosing the
initial mass of fluid which yields the smallest $P_c$, {\it i.e.} the film
thickness where the derivative $-\partial_{hh} f(h,x)|_{\bar{h}}$ is maximal
(for constant $G$), see also  \cite{KKS00b}.
The transition from coarsening to pinning is hysteretic, giving rise 
to an extended region where the desired pinned pattern coexists with 
coarser structures. 
Our results and the methodology also apply to other strategies of
stabilizing a periodic pattern including introduction of anisotropy or
convective flows \cite{GNDZ01}.

%

\begin{thebibliography}{10}
\bibitem{noteallsol}
 We computed all stationary solutions with $P=n P_{het}$ and $n~\le~10$ 
 but the qualitative picture is not changed.
\bibitem{stab}
 Space has been discretized by 64 Fourier modes and the spectrum of eigenvalues
  as well as the corresponding eigenmodes are calculated from the linearization
  of Eq.~(\ref{h4}) around the stationary solutions.
\end{thebibliography}
%

\end{document}